\documentclass[journal=acscii,manuscript=article]{achemso}
\setkeys{acs}{etalmode=truncate,maxauthors=0}

\usepackage[version=3]{mhchem} 
\usepackage{amssymb}
\usepackage{amsfonts}
\usepackage{amsmath}
\usepackage{multirow}
\usepackage{epsfig}
\usepackage{hyperref}

\usepackage{soul}
\usepackage[usenames,dvipsnames]{xcolor}
\usepackage[utf8]{inputenc}
\usepackage{array}
\usepackage{wrapfig}
\usepackage{multirow}
\usepackage{tabularx}

\usepackage{titletoc}
\titlecontents{figure}[0mm]%
    {\makebox{Figure~}}%
    {\makebox{\thecontentslabel: }}%
    {}%
    {\enspace\dotfill\enspace\thecontentspage}

\author{João N. C. Especial} 
\affiliation{BioISI – Instituto de Biossistemas e Ci{\^e}ncias Integrativas and Departamento de Física, Faculdade de Ci{\^e}ncias, Universidade de Lisboa, 1749-016, Lisboa, Portugal}

\author{Beatriz P. Teixeira} 
\affiliation{Departamento de Física, Faculdade de Ci{\^e}ncias, Universidade de Lisboa, 1749-016, Lisboa, Portugal}

\author{Ana Nunes} 
\affiliation{BioISI – Instituto de Biossistemas e Ci{\^e}ncias Integrativas and Departamento de Física, Faculdade de Ci{\^e}ncias, Universidade de Lisboa, 1749-016, Lisboa, Portugal}

\author{Miguel Machuqueiro} 
\affiliation{BioISI – Instituto de Biossistemas e Ci{\^e}ncias Integrativas and Departamento de Química e Bioquímica, Faculdade de Ci{\^e}ncias, Universidade de Lisboa, 1749-016, Lisboa, Portugal}

\author{Patrícia F. N. Faísca} 
\email{pffaisca@ciencias.ulisboa.pt}
\affiliation{BioISI – Instituto de Biossistemas e Ci{\^e}ncias Integrativas and Departamento de Física, Faculdade de Ci{\^e}ncias, Universidade de Lisboa, 1749-016, Lisboa, Portugal}

\title{The role of topology on protein thermal stability}

\begin{document}

\begin{abstract}
For several decades, experimental and computational studies have been used to investigate the potential functional role of knots in protein structures. A property that has attracted considerable attention is thermal stability, i.e., the extent to which a protein retains its native conformation and biological activity at high temperatures, without undergoing denaturation or aggregation. Thermal stability is quantified by the melting temperature $T_m$, an equilibrium property that corresponds to the peak of heat capacity in differential scanning calorimetry (DSC) experiments.
Experimental and computational studies report conflicting effects of knotting on protein thermal stability \cite{Soler2013, Danny2019b}. Here, we use extensive Monte Carlo simulations of a simple C-alpha model of protein YibK, with energetics modeled by the Go potential, to show that $T_m$ does not  depend on the topological state of the protein. Our simulations further support the view that the discrepancy between the experimental and computational results stems from a pronounced separation of timescales for unknotting and unfolding that is inherent to deeply knotted proteins like YibK. In particular, the timescale separation implies that the complete unfolding–untying transition may not be accessible within the duration of a DSC experiment, whose apparent $T_m$ measurements likely reflect a non-equilibrium distribution lacking unfolded states that are also unknotted.
\end{abstract}

\section{Introduction}
\label{sec:intro}
A knotted protein is a protein whose polypeptide chain is self-tied, forming a physical (i.e., open) knot in its native structure. Although the first knotted protein was reported in 1977~\cite{richardson1977}, it was not until 2000 that Taylor's computational methodology ~\cite{taylor2000}  (which enables a systematic determination of the topological state of open polymer chains) brought these complex proteins into the spotlight. Knotted proteins make up 1\% of all protein entries (225 k) in the Protein Data Bank (PDB)~\cite{Sulkowska2018}, and are statistically rare compared to compact random loops of the same size~\cite{Grosberg2006}. All three domains of life contain knotted proteins~\cite{Virnau2006}, including Archaea, a group of single-celled organisms believed to have inhabited the planet for around 3.8 billion years. A recent survey that analyzed 700 k knotted proteins
predicted by the computer artificial intelligence system AlphaFold (AF)~\cite{AlphaFold}, and 42 proteomes (16 bacteria, 9 fungi, 4 plants, and 13 animals), reported that the percentage of knotted proteins in any given proteome is around 0.4\%~\cite{Sulkowska2024}. \par

The minimum number of crossings in all possible planar projections of a given chain is an incomplete topological invariant used to classify both topological knots (which are closed curves in space) and protein knots. The $3_1$ (or trefoil) knots, a class that features three crossings, is by far the most prevalent protein knot, followed by the $4_1$ and $5_2$ knots, with four and five crossings, respectively~\cite{Sulkowska2018, Sulkowska2024}. The most complex protein knot has seven crossings (it is a $7_1$ knot) and was predicted by AF~\cite{AlphaFold}. It is an 89-residue-long bacterial protein (termed Q$_9$PR$_{55}$) whose atomic structure was recently experimentally resolved, validating AF's predictions~\cite{Danny7_1}. \par 
In the context of structural biology, the design principle 'form follows function' means that a protein's structure is adapted to successfully perform its specific task. 
It highlights the fact that protein structure is not arbitrary but is shaped by the specificities of its role, which ultimately contributes to the organism's survival.  
The fact that knotted proteins are statistically rare does not imply a lack of function for the knot; rather, it indicates that folding into a knotted native structure is costly (e.g, it leads to slow folding and topological frustration \cite{Yeates2010}) and therefore unlikely to be tolerated unless compensated by specific advantages. If knotting were purely neutral or deleterious, one would expect it to be entirely suppressed by evolution. The fact that knotted proteins have been conserved throughout evolution~\cite{Joanna2012} (and, in many cases, have likely evolved from knotted precursor structures in organisms dating back billions of years \cite{Tubiana2024}) suggests that they may confer some advantage to their carriers. Thus, it is not surprising that a large body of theoretical and experimental work has been devoted to determining the potential functional roles of knots in proteins (reviewed in~\cite{Jackson2020, Tubiana2024}), alongside establishing how they fold into their intricate native structures (reviewed in ~\cite{Faisca2015, Jackson2017, Danny2023, Tubiana2024}). While significant advancements have been made regarding folding and knotting mechanisms, a consensus has not been reached concerning the function of protein knots \cite{Jackson2020}. Knots can potentially influence proteins at both the level of physical properties, structure, as well as at the level of their biological function. A recent study found that the $5_2$ knot in protein UCHL-1 is critical for the preservation of secondary structure~\cite{Sara2024}. Previous studies have reported that a knotted structure helps shape and form the binding site of enzymes~\cite{Nureki2002, Nureki2004, Jacobs2002}. Additionally, one study has shown that the knotted region is directly involved in enzymatic activity~\cite{Ya-Ming2016}. Indeed, for the protein TrmD (an example of an important class of enzymes, the SPOUT and SAM synthases), it was shown that the knot regulates catalysis, being critical to protein function~\cite{Ya-Ming2016}. Examples of physical properties are thermal stability, which measures a protein's resistance to melting and is quantified by the melting temperature; kinetic stability, which measures the protein's resistance to unfolding at a certain temperature and is quantified by the unfolding rate constant; and mechanical stability, a measure of the protein's ability to withstand physical forces and deformations without unfolding or losing its native structure. Simulation studies using models with different levels of resolution suggested that knots in proteins enhance kinetic stability~\cite{Soler2014, Joanna2008}, while an enhancement of mechanical stability has been reported for several model systems based on {\it in vitro} experiments~\cite{Danny2020, Danny7_1,Danny2023}) and computational studies~\cite{Joanna2008, Xu2018}. \par

Some years ago, we used Monte Carlo simulations of a simple lattice model to find that a knotted topology does not affect protein's thermal stability~\cite{Soler2013}. However, this contradicts a recent experimental study that reported a significant enhancement in the thermal stability of protein YibK, which features a deep trefoil knot~\cite{Danny2019a}. 
In lattice models, the discretization of conformational space restricts backbone geometry, hindering the study of specific proteins like YibK, which is only possible in the context of a continuum off-lattice model. Additionally, the reduced conformational freedom inherent to lattice models means that they cannot accurately capture entropic effects.  Thus, the purpose of the present work is to revisit the question of whether or not thermal stability is affected by the presence of an open knot in the native structure of protein YibK. Our results strongly support the view that knots play no role in enhancing protein thermal stability or any other thermodynamic equilibrium property.

\section{Model and Methods}
\label{sec:methods}
\subsection{The C$_\alpha$ G\={o} model}
\begin{figure}[hbt!]
    \centering
   \includegraphics[width=0.9\linewidth]{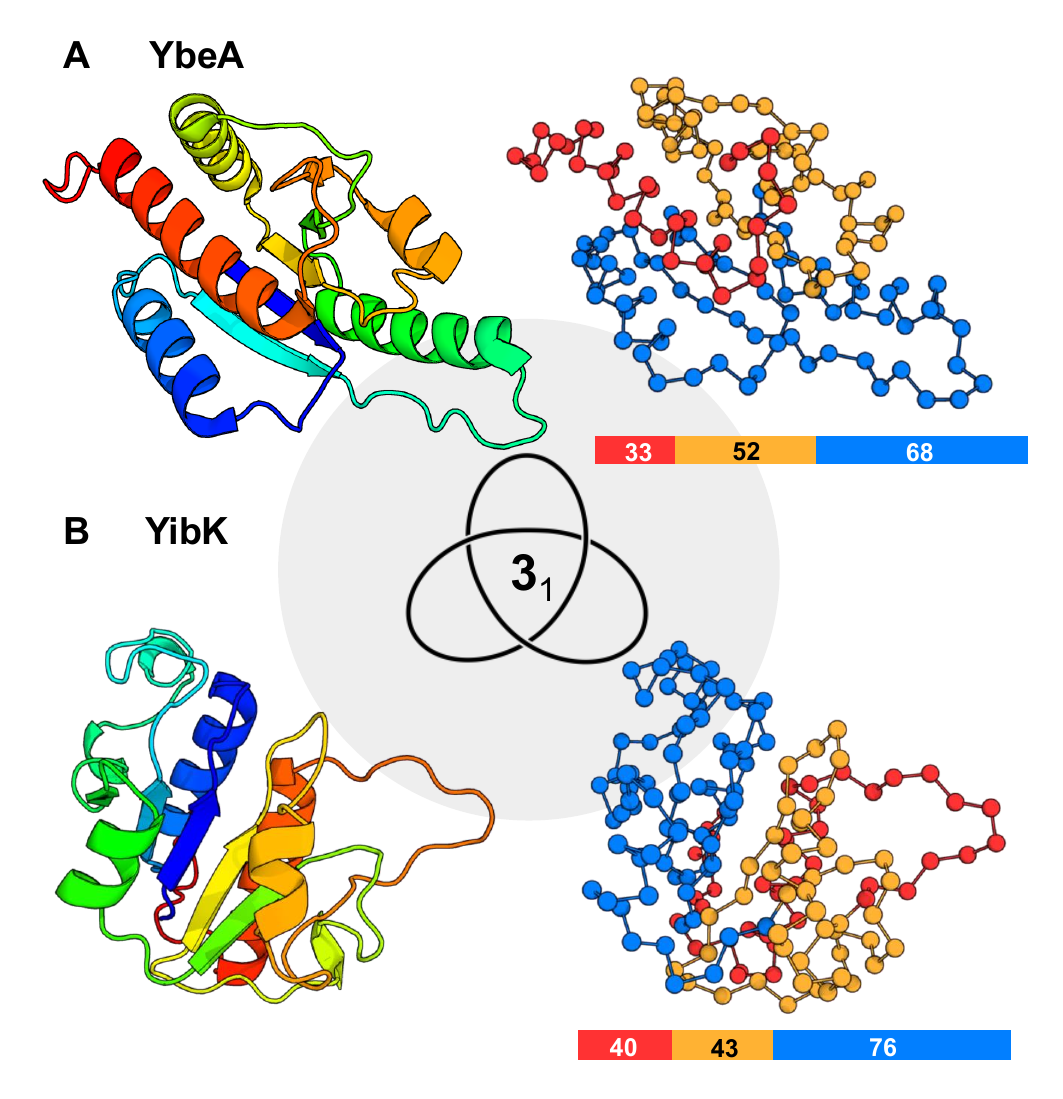}
  \caption{Model systems used in the present study. Cartoon representation (left) and bead and stick representation (right) of the native structure of protein YbeA (PDB id: 1ns5, chain A) (A), and YibK (PDB id: 1mxi, chain A) (B) with the knotted core colored in orange and the knot tails (number of residues that must be removed from the chain termini to untie the knot) colored in red (C-terminus) and blue (N-terminus). Each bead represents a C$_\alpha$ atom and rigid sticks represent pseudo-bonds connecting pairs of C$_\alpha$ atoms. The size (measured in number of beads) of the knot tails and knotted core is indicated. In both proteins, the knotted topology results from threading the C-terminus through the knotted core.}
    \label{fig:native_structure}
\end{figure}

Proteins are represented by a simple C$_\alpha$ model (Figure \ref{fig:native_structure}). Accordingly, residues are reduced to hard spherical beads of uniform size, centered on the C$_\alpha$ atoms. Consecutive C$_\alpha$ atoms are connected by rigid sticks representing pseudobonds on the amide planes. We adopt a radius of 1.7~\AA{} for the beads, which is the van der Waals radius of C$_\alpha$ atoms \cite{Tsai1999}. For the length of each stick, we adopt the distance between the C$_\alpha$ atoms of the respective bonded residues in the protein's native conformation, these being approximately 2.9~\AA{} for cis bonds and 3.8~\AA{} for trans bonds.
Two non-bonded residues are said to be in contact in the native conformation if the smallest distance between any two heavy atoms, one belonging to each residue, is $\leqslant$ 4.5 \AA, this cut-off being chosen because it is slightly larger than twice the average van der Waals radius of heavy atoms in proteins.

To model protein energetics, we consider the native-centric G\={o} potential \cite{Go}. Accordingly, the total energy $E$ of a conformation defined by bead coordinates $\{\vec{r}_i\}$ is given by
\begin{equation}
    E\left(\{\vec{r}_i\}\right) = \varepsilon \sum_{i,j \geqslant i+2}^N
    \left[\left(\frac{\left|\vec{r}_i - \vec{r}_j\right| - \left|\vec{r}_i^{\, nat} - \vec{r}_j^{\, nat}\right|}{w}\right)^2 +
    1\right]^{-1}
    \left(\chi_{ij} \chi_{ij}^{nat} + \chi_{ji} \chi_{ji}^{nat} + \frac{1}{2}\right)
    \Delta_{ij}^{nat} \, .
\end{equation}
where $N$ is the chain length measured in the number of beads, $\vec r_{i}^{nat}$ is the position vector of bead $i$ in the native structure, $\Delta_{ij}^{nat}$ is 1 if the $i-j$ contact is present in the native conformation and is 0 otherwise, $\varepsilon$ is a uniform intramolecular energy parameter (taken as $-1$ in this study, in which energies and temperatures are shown in reduced units), $w$ is the half-width of the inverse quadratic potential well, and the chirality of contact $i-j$ in the conformation under consideration is
\begin{equation}
    \chi_{ij} = \Theta\left( \, \left(\vec{r}_i - \vec{r}_j\right)^{ } \cdot
                [ (\vec{r}_{j+1} - \vec{r}_j) \times
                (\vec{r}_{j-1} - \vec{r}_j) ] \, \right) - \frac{1}{2} \, .
\end{equation}

The chirality of the $i-j$ contact in the native conformation is
\begin{equation}
    \chi_{ij}^{nat} = \Theta\left( \, \left(\vec{r}_i^{\, nat} - \vec{r}_j^{\, nat}\right)^{ } \cdot
                [ (\vec{r}_{j+1}^{\, nat} - \vec{r}_j^{\, nat}) \times
                (\vec{r}_{j-1}^{\, nat} - \vec{r}_j^{\, nat}) ] \, \right) - \frac{1}{2} \, .
\end{equation}
In equations (2) and (3), $\Theta$ is Heaviside's unit step function, which takes the value 1 if its argument is greater than zero and the value 0 otherwise.
The chirality factor in (1) favors the native conformation \emph{vis à vis} its mirror conformation. A native contact is considered formed if the distance between the centers of the respective beads differs from the distance between their C$_\alpha$ atoms in the native conformation by less than the half-width of the potential wells, $w$.

\subsection{Replica-exchange Monte Carlo simulations}
The conformational space of the C-alpha model is explored with Metropolis Monte Carlo (MC) \cite{Metropolis(1953)} by means of a move set that comprises crankshaft and pivot moves. All simulations start from an unfolded (and unknotted) conformation obtained from a high-temperature simulation, and folding progress is monitored using the fraction of native contacts, $Q$. To sample equilibrium distributions, we use Monte Carlo replica-exchange (MC-RE) \cite{Sugita(1999)}. 

To explore the conformational space, two sampling schemes can be deployed. A canonical one in which the move set preserves the linear topology of the chain, and another one that breaks the linear topology of the chain by allowing it to cross itself~\cite{Especial2022}, while preserving the sequential bonding of the protein chain and excluded volume among its units. We have previously shown, using simple C$_\alpha$ protein models (of unknotted proteins Fn-III and $\beta_2$m, shallowly knotted protein MJ0366, and deeply knotted protein Rds3p), that there is no difference in equilibrium properties calculated with a move set that preserves linear topology and one that does not~\cite{Especial2022}. However, the move set that breaks the linear topology of the chain is extraordinarily advantageous in the case of deeply knotted proteins like the ones studied here, since it provides correct equilibrium results at a much lower computational cost~\cite{Especial2022}. Thus, unless otherwise stated, we will deploy the sampling method that does not preserve the linear topology of the chain.

The weighted histogram analysis method (WHAM) \cite{Chodera(2007)} is used to analyze data from the MC-RE simulations and to produce maximum-likelihood estimates of the density of states, from which expected values for thermodynamic properties are calculated as functions of temperature. An example is the heat capacity, $C_V$, defined in reduced units as $C_V = (<E^2> - <E>^2) / T^2$. The melting temperature, $T_m$, is determined as the temperature at which the $C_V$ peaks. 
We measure the cooperativity degree of the transition by the ratio of the full width at half maximum ({\it FWHM}) of the $C_V$ peak to the melting temperature, and the half-width of the potential well, $w$, is adjusted to obtain a simulated {\it FWHM}$/T_m$ ratio between 4 and 5\% \cite{Larriva2010, Fernandez2021}. This criterion has been successfully used in previous simulations employing a similar potential and sampling \cite{faisca2016, faisca2019}. WHAM is also used to project the density of states along $Q$ to obtain free energy profiles at a specific temperature. \par
 
\subsection{Molecular Dynamics simulations} 
The MD simulations were performed using GROMACS~2024.3 \cite{abraham2015}, with the AMBER~14SB force fields \cite{maier2015}, and the TIP3P water model \cite{jorgensen1983, neria1996}. The system consisted of a protein, 8808~water molecules, and 4~Cl$^-$ ions to equilibrate the overall system charges. We performed a two-step minimization procedure using the steepest descent method and an initialization protocol with two additional steps, in which temperature (NVT) and pressure (NPT) baths are turned on and equilibrated for 100~ps segments, with an integrator step of 1~fs. The nonbonded interactions were computed using a single cutoff of 1.0~nm, with the neighbor list updated every 20~steps. Van der Waals interactions were truncated beyond this cutoff, while long-range electrostatics were calculated using the Particle-Mesh Ewald (PME) method \cite{darden1993}, with a Verlet cutoff of 1.0~nm and a Fourier grid spacing of 0.125~nm. Bond lengths involving hydrogen atoms in solute and water molecules were constrained using the LINCS \cite{hess2008b} and SETTLE \cite{miyamoto1992} algorithms, respectively. In the NVT ensemble, the v-rescale thermostat \cite{bussi2007} was employed to maintain a temperature of 310~K (with a coupling constant of 0.1~ps). In the NPT ensemble, the v-rescale thermostat \cite{bussi2007} (coupling constant of 1.0~ps) was used in combination with the Parrinello-Rahman barostat \cite{parrinello1981} (coupling constant of 2.0 ps and isothermal compressibility of 4.5$\times$10$^{-5}$~bar$^{-1}$). For the MD relaxation step, we run for 100~ns with the integration step adjusted to 2~fs. 
 
\subsection{Topological state}
The topological state (knotted or unknotted) of a sampled conformation is determined using the Koniaris-Muthukumar-Taylor (KMT) algorithm \cite{taylor2000}.

\section{Results and discussion}
\label{sec:results}

\subsection{Model systems}
\label{sec:systems}
Proteins YbeA (from {\it Escherichia coli}, PDB ID: 1ns5) and YibK (from {\it Haemophilus influenzae}, PDB ID: 1mxi and PDB ID:1j85) are homodimeric enzymes that belong to the $\alpha$\textbackslash$\beta$-knot superfamily of MTases. In this work, we consider the monomer identified as chain A in the corresponding PDB files. The native structure of both proteins features a trefoil knot. In a knotted protein, the region of the polypeptide chain encompassing the knot is termed the knotted core. Protein knots are classified as deep or shallow, depending on the number of residues that must be removed from one of the termini to unknot the chain. If the latter is larger (or smaller) than 20, the knot is deep (or shallow)\cite{Taylor2003, Jackson2017, Joanna_Israelli}. YbeA has a chain length of 155 amino acids, and its knotted core extends from residue 69 to residue 121. Consequently, it is necessary to remove 68 residues from the N-terminus (or 34 residues from the C-terminus) to untie the knot that is classified as deep. The native structure of YibK also embeds a deep trefoil knot. Indeed, YibK has a chain length of 160 residues and a knotted core that extends from residue 77 to 120, requiring removal of 76 residues from the N-terminus (or 40 residues from the C-terminus) to untie the knot. Despite their structural similarity, these bacterial proteins share only 19\% of sequence identity.

\begin{figure}[hbt!]
    \centering
   \includegraphics[width=0.99\linewidth]{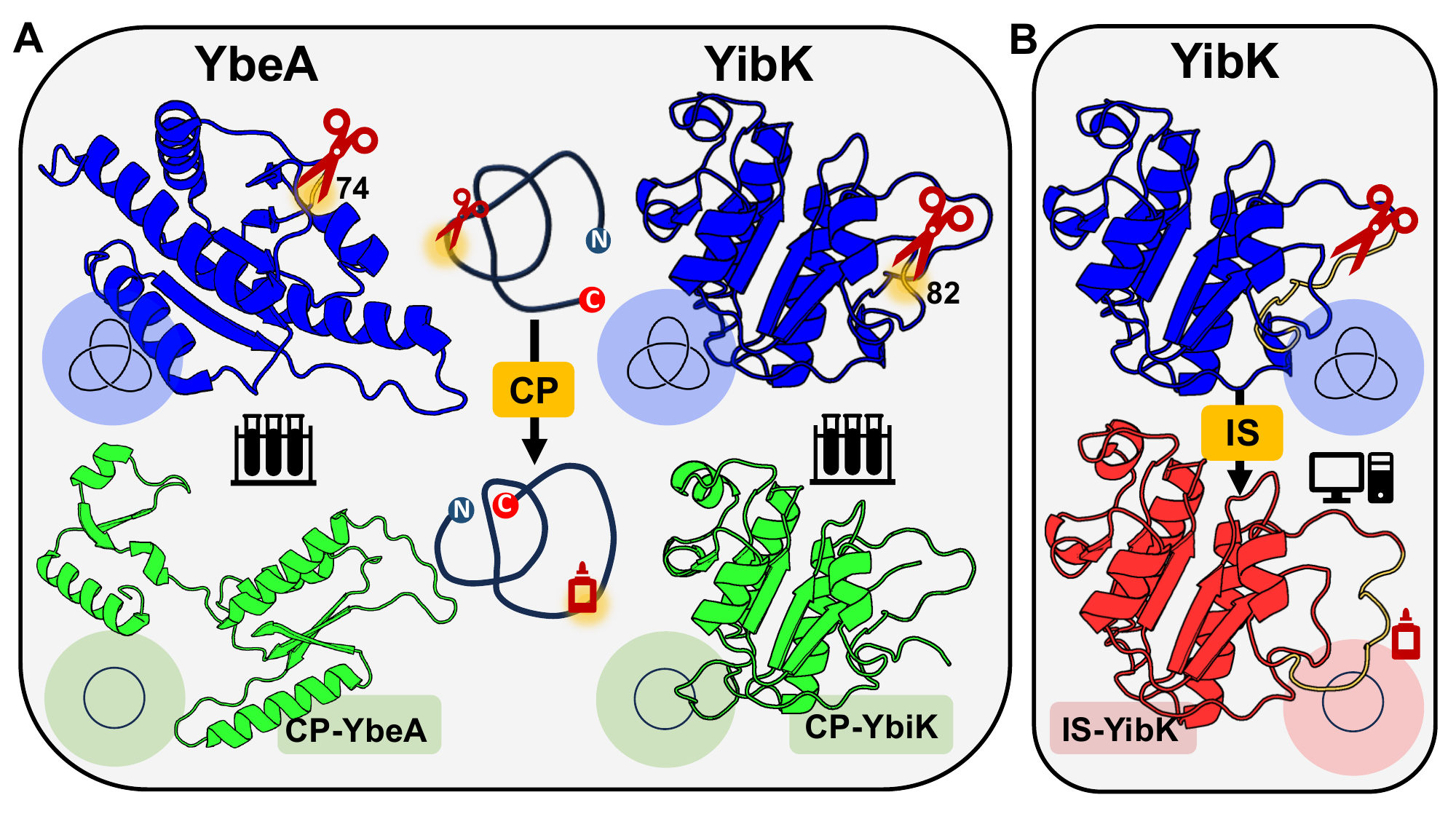}
  \caption{Control systems used in the present study. Unknotted control systems CP-YbeA (PDB ID: 1ns5) and CP-YibK (PDB ID: 6ahw) obtained experimentally by Hsu and co-workers through circular permutation of YbeA (PDB ID: 1ns5) and YibK (PDB ID: 1mxi), respectively (A), and \textit{in silico} unknotted control system IS-YibK, obtained through computer modeling from YibK (PDB ID: 1mxi)(B). The knotted structures are all shown in blue, the unknotted structures obtained experimentally by CP are shown in green, and the unknotted structure obtained \textit{in silico} is shown in red.}
  \label{fig:control_systems}
\end{figure}

To explore the role of the knotted topology in determining equilibrium properties, it is necessary to consider control systems~\cite{Yeates2010}. A suitable control system is a similar, yet unknotted protein that must be used for comparison. Since knotted/unknotted pairs do not exist in nature, they must be designed or engineered. In the present study, we consider two control systems obtained experimentally and one control system created through computational modeling. The {\it in vitro} control systems were engineered by Hsu and co-workers through a process known as circular permutation~\cite{Danny2019a, Danny2019b}. A circular permutant (CP) is an engineered protein that results from linking the C- and N-terminus of the polypeptide chain after disrupting the protein's backbone at some selected site~\cite{Faisca2012}. In the case of protein YbeA, an unknotted control system was obtained by creating a CP after breaking the chain at the knotted core in position 74,
while for YibK
the chain was broken at position 82, also in the knotted core, to create the CP (Figure \ref{fig:control_systems}A). In the present study, the resulting unknotted structures are designated by CP-YbeA (PDB ID: 5zyo) and CP-YibK (PDB ID: 6ahw). The {\it in silico} control structure was obtained for YibK (PDB ID: 1mxi) by manually passing the loop containing residues 114--125 across the knotted core (amino acids 81-85) (Figure \ref{fig:control_systems}B). The manipulated structure was then subjected to an MD simulation, in which segments 1-80, 86-113, and 126-156 were positionally restrained 1000~kJ/mol~nm$^2$), while the two loops (81--85 and 114--125) were allowed to relax. The final relaxed and refined structure is named IS-YibK.

For all  structures, we computed the gyration radius ($R_\mathrm{g}$), which is a measure of structural compactness, the root-mean-square-deviation (RMSD) to the wild-type native structure, which is a measure of structural similarity, and the number of native contacts, which is a measure of the system's energy in the context of the adopted model. These properties are reported in Table \ref{Tbl:properties}. The native structure of CP-YbeA exhibits significant structural rearrangements, which are visible in the cartoon representation (Figure \ref{fig:control_systems}A), and results in a substantially high RMSD ($>>$3\AA). There is also a significant loss of compactness, resulting in a high gyration radius and a substantial loss ($\sim$25\%) of native contacts. On the other hand, both the unknotted CP-YibK and IS-YibK share a high degree of structural similarity with YibK (RMSD$<<$2\AA). However, while IS-YibK has fewer native contacts than YibK ($\sim$11\%), the circular permutant exhibits a nearly identical degree of compactness and essentially the same number of native contacts (the difference between the two model systems being only 2\%), which makes it the best control system considered here.

For each knotted protein and corresponding control system, we measured several equilibrium properties as a function of temperature. 

\begin{table}[hbt!]
\caption{Properties of the model proteins considered in this study, where K stands for knotted, and U stands for unknotted. The RMSD of the unknotted controls is measured against the corresponding knotted structure.}
\label{Tbl:properties}
\begin{center}
\begin{tabular}{r|cccc}
\hline
Protein & Topological state & RMSD (\AA) & $R_\mathrm{g}$ (\AA) & No. native contacts \\
\hline
YbeA & K & - & 15.5 & 542 \\
CP-YbeA & U & 16.6 & 21.8 & 410 \\
\hline
YbiK & K & - & 15.9 & 552 \\
CP-YibK & U & 0.35 & 15.5 & 561 \\
IS-YibK & U & 0.48 & 16.6 & 500 \\
\hline
\end{tabular}
\end{center}
\end{table}



\subsection{Protein YbeA and its control system}
\begin{figure}[hbt!]
    \centering
   \includegraphics[width=1.0\linewidth]{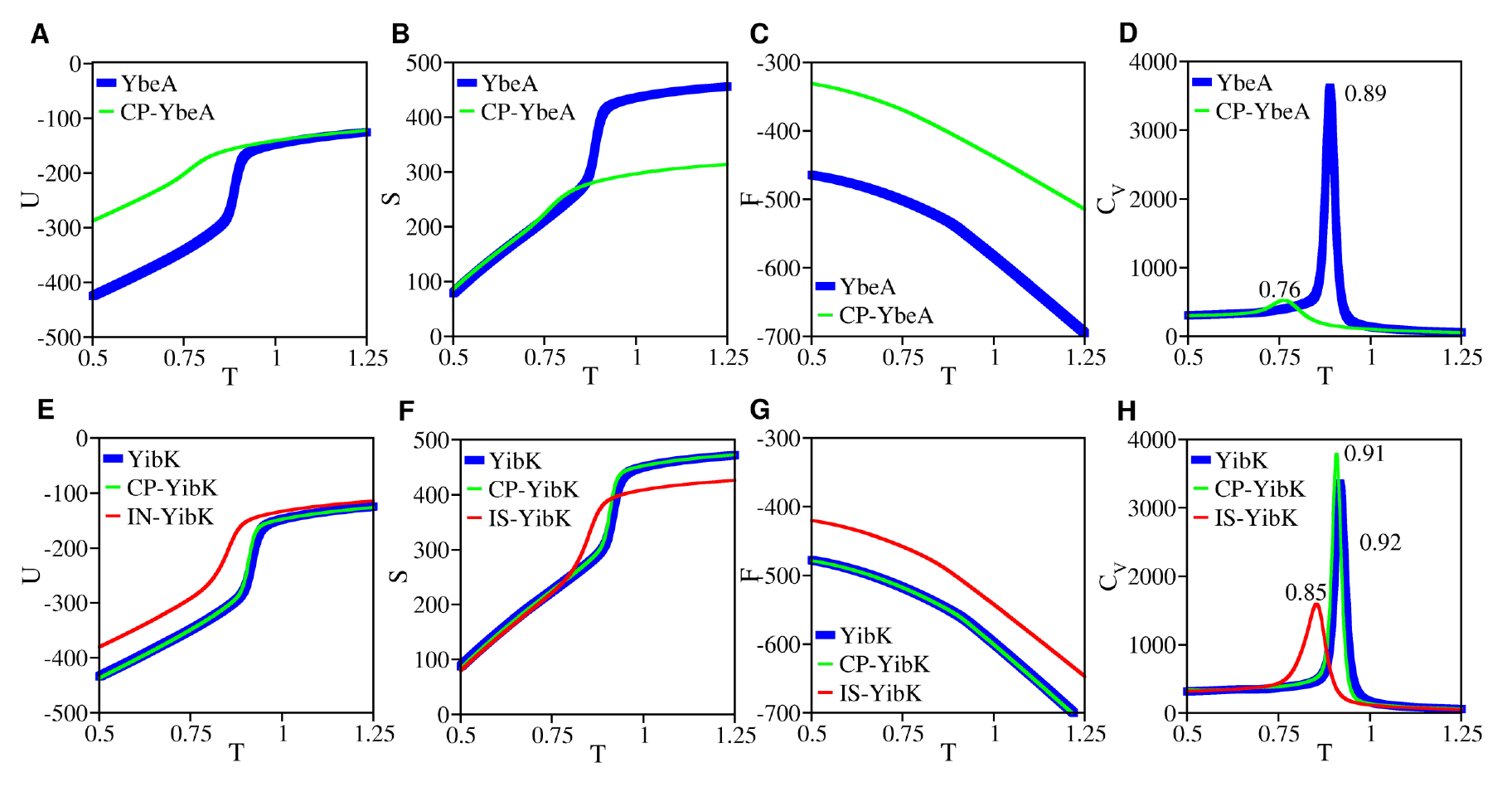}
  \caption{Equilibrium properties as a function of temperature obtained from Monte Carlo simulations. Dependence of the internal energy ($U$), entropy ($S$), free energy ($F$), and heat capacity ($C_V$) on temperature ($T$) for the knotted (K) and unknotted (U) model systems of YbeA (A-D) and YibK (E-H). The melting temperature, $T_m$, is indicated on the corresponding heat capacity curves. }
    \label{fig:thermo}
\end{figure}

\begin{figure}[hbt!]
    \centering
   \includegraphics[width=0.7\linewidth]{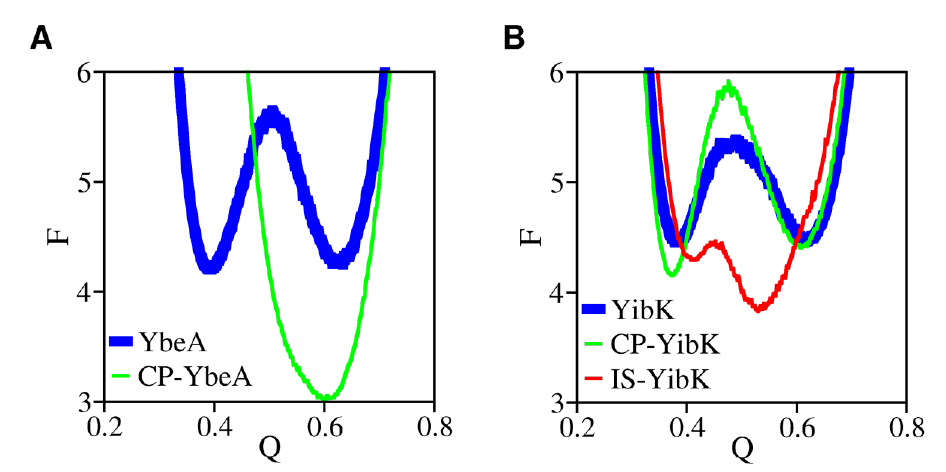}
  \caption{Free energy profiles obtained from Monte Carlo simulations. Dependence of the free energy ($F$) on the fraction of native contacts $Q$ at the melting temperature ($T_m$) for the YbeA (A) and YibK (B) model systems.}
    \label{fig:thermoQ}
\end{figure}

Results reported in Figure~\ref{fig:thermo}A--D show that YbeA folds cooperatively in a two-state manner, while the control CP-YbeA folds non-cooperatively. This lack of cooperativity is particularly evident from the rather gentle slope of the internal energy curve (Figure \ref{fig:thermo}A), and from the broadness of the heat capacity curve (Figure \ref{fig:thermo}D). In addition, the melting temperature of CP-YbeA is significantly ($\sim$15\%) lower than that of YbeA. 
We have also measured the dependence of the free energy on the fraction of native contacts ($Q$) at the melting temperature for the knotted-unknotted pair. We find that the behavior of the folding transition changes from two-state (for YbeA) to downhill (or barrierless) (for CP-YbeA) (Figure \ref{fig:thermoQ}A). These findings, in particular the decreased thermal stability of CP-YbeA, cannot be attributed solely to its unknotted topological state. Indeed, untying the knot in YbeA introduced drastic structural changes and a significant loss of native interactions, which collectively account for the observed behavior.

\subsection{Protein YibK and its control systems}
To isolate the role of the topological state on thermal equilibrium properties, particularly thermal stability, we now examine YibK and its control systems. In this case, the two-state nature of the folding transition is conserved across the three systems (Figure~\ref{fig:thermo}E-H). A minor, yet non-negligible difference is observed between YibK and its \textit{in silico} control system, IS-YibK. In particular, for the latter, the melting temperature is 7\% lower than that of YibK (Figure~\ref{fig:thermo}H). Since the number of native contacts is not conserved in the unknotted protein, we cannot ascribe its lower thermal stability to its unknotted topological state. In fact, when one compares YibK and CP-YibK, which are nearly structurally identical and share almost the same number of native contacts, their thermodynamic behavior is practically identical, with the thermal stability differing by only 1\%. In line with these findings, the similarity between the free energy profiles (Figure~\ref{fig:thermoQ}B) is also significantly higher for YibK and its unknotted circular permutant.

\subsection{Comparison with experimental data}

The results reported in the previous section show that as the similarity between the structures of a knotted protein and its unknotted control increases, differences in thermodynamic equilibrium behavior diminish. Indeed, simulations indicate that an unknotted control preserving the same native geometry (and, consequently, distances between atoms) would exhibit similar thermodynamic behavior. From a physics perspective, this is expected: in our model, the Hamiltonian depends only on the native contact map (atomic positions and corresponding interactions) and is independent of the topological state. All native interactions are identical between the knotted and unknotted structures, and non-native interactions are neutral by construction.  Therefore, if the native contact maps of knotted and unknotted structures are identical, (or nearly identical as in the case of YibK and its control system CP-YibK) an identical (or nearly identical) density of states is expected, and, consequently an identical (or nearly identical) thermodynamic behavior as well.

However, the experimental measurements carried out for YibK and CP-YibK do not support the findings from simulations, indicating that the knotted topology enhances the protein's thermal stability. Indeed, a decrease in the melting temperature of 22\%, measured with differential scanning calorimetry (DSC), or 27\%, measured with far-UV circular dichroism (CD) signals, was reported~\cite{Danny2019a} for the unknotted protein. 

Part of the decrease in thermal stability observed experimentally may be due to a 13\% reduction in the dimer interface per chain in the dimeric version of CP-YibK. 
A fraction of the decreased stability could then be attributed to the weakened dimer interface, featuring fewer intermolecular hydrogen bonds and salt bridges \cite{Danny2019a}.  Because experimental measurements were performed on the dimer while simulations focused on the monomer, a direct comparison between experimental and computational data is not straightforward. Furthermore, the thermal unfolding experiments were not demonstrated to be reversible, unlike the simulations. If unfolding is irreversible, the reported experimental melting temperatures reflect apparent rather than true thermal stability. If the control system CP-YibK used in experiments is more prone to aggregation than the knotted YibK, this could also easily suppress the experimental melting temperature. Although simulations provide a controlled framework to isolate the role of protein topology in thermal stability, they rely on approximations at both the level of protein representation and the Hamiltonian. Ultimately, this means that the simulation results should be interpreted with care rather than providing definitive thermodynamic measurements.  Despite the limitations inherent in both experimental and computational approaches, their results provide different views of the role of topology in determining YibK’s thermal stability. In the following, we discuss a possible, fundamental reason for this discrepancy.

A few years ago, Jennings et al. \cite{Jennings2016} reported the existence of two distinct timescales for unfolding and untying, upon chemical denaturation, in the deeply knotted protein MTTTm (PDB ID: 106d), which belongs to the same family as YibK and has a comparable chain length. They found that the transition to the untied states required at least six months, whereas unfolding occurred on a timescale of approximately two weeks. A previous study by Jackson and co-workers reported that the knotted topology is retained in the unfolded state of YibK, following chemical denaturation \cite{Jackson2010}. These experimental observations are consistent with insights from computational simulations based on lattice models that also reported the occurrence of knotted conformations in thermally denatured states of deeply knotted lattice proteins.\cite{Soler2013}. 

\begin{figure}[hbt!]
    \centering
   \includegraphics[width=0.9\linewidth]{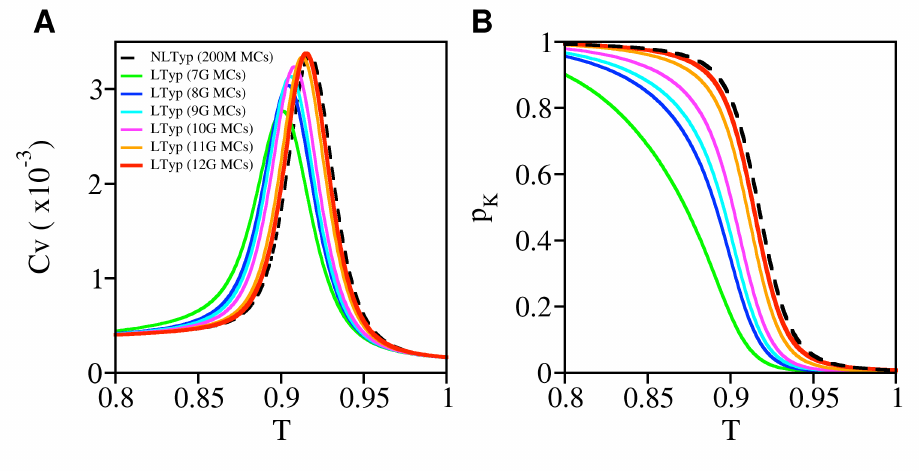}
  \caption{Depedence of the heat capacity (A) and of the knotting probability (B) on the simulation temperature for YibK. The MC simulation that does not preserve the linear topology of the chain (NLTyp)  requires 200$\times 10^6$ MC steps to equilibrate (the curves  do not change for a larger number of MC steps). On the other hand, a simulation that preserves the linear topology of the chain requires 12$\times 10^9$ MC steps to  equilibrate. Interestingly, a non-equilibrated simulation predicts an apparent $T_m$ that is smaller, and a transition that is clearly less sharp (as denoted by the shape of the knotting probability curve) than that predicted by the equilibrated sampling. }
    \label{fig:equi_time}
\end{figure}

As discussed in the Model and Methods section, achieving equilibrium distributions for complex polymeric systems, such as deeply knotted proteins, is particularly challenging when using off-lattice Monte Carlo (MC) simulations with a move set that preserves the chain's linear topology, starting from unfolded, unknotted conformations. For instance, in simulations of the deeply knotted protein Rds3p, which is 51~amino acids shorter than YibK, the linear topology-preserving move set required an order of magnitude more MC steps to reach equilibrium compared to simulations that allow chain crossings \cite{Especial2022}. By contrast, for unknotted proteins, such as Fn-III and $\beta$2-microglobulin, the computational cost is independent of the move set employed \cite{Especial2022}. The fact that, for unknotted proteins, the equilibration time is independent of the move set, whereas deeply knotted proteins exhibit a strong dependence, indicates that knotting introduces a significant degree of topological frustration, which delays the equilibration time,  when chain crossings are forbidden. We repeated the analysis of the equilibration time for YibK and found that enforcing linear topology preservation increases the equilibration time by 1.78 orders of magnitude relative to simulations that do not preserve the linear topology of the chain (Figure~\ref{fig:equi_time}).    Since real protein dynamics also preserves the linear topology of the polypeptide chain, it is plausible that the scan rate (i.e., heating rate) used in the DSC experiments is too fast for the knotted protein to equilibrate at each temperature, and even more so to allow the population of denatured conformations that are also unknotted. Thus, the pronounced separation of timescales for unknotting and unfolding suggests that, for deeply knotted proteins such as YibK, the complete unfolding–untying transition may not be accessible within the time frame of a DSC experiment. In this case, the experimental measurements likely reflect a non-equilibrium distribution lacking unfolded states that are also unknotted; in that case, the measured melting temperature represents an apparent value rather than a true thermodynamic parameter.

\section{Conclusions}
Over the past 20 years, a remarkably large body of theoretical and experimental work has been dedicated to the study of knotted proteins\cite{Faisca2015, Jackson2017, Danny2023}. A major line of research has focused on whether knots confer any functional advantage to their carriers \cite{Jackson2020}. Answering this question will provide insight into the evolutionary significance of knotted proteins and may have an impact on the design of drugs for technological or medical applications.\par 

A major challenge in studying the role of knots in proteins is the construction of a suitable control system, a version of the protein that is nearly identical to the knotted variant but lacks the knot \cite{Yeates2010}. In the context of lattice models, it is quite straightforward to untie a model protein while introducing minimal changes to the structure \cite{Faisca2010}, and previous simulation efforts in the lattice context have shown that a knotted topology does not enhance the protein's thermal stability \cite{Soler2013}. Unfortunately, in more realistic off-lattice models, as well as in real-world proteins, engineering unknotted control systems is not as straightforward.\par
Here, we conducted extensive Monte Carlo simulations of a simple off-lattice model with energetics governed by a Go potential to show that protein knots can strongly affect the equilibration time and make the equilibrium distribution not accessible in a reasonable timescale. Indeed, by considering the chain's linear topology as either `on’ or `off’, our simulations corroborated that knots impose this limitation, with the move set that preserves the linear topology requiring almost two orders of magnitude more Monte Carlo steps to equilibrate. Additionally, by comparing the knotted protein YibK with a nearly identical control system experimentally engineered \cite{Danny2019a}, our results show that, if equilibrium is indeed reached, knotting does not enhance the thermal stability of this deeply knotted protein. This finding contradicts experimental data analysing the same pair of structures, which reported a moderate decrease in thermal stability for the unknotted one \cite{Danny2019a}.  
Based on results from the simulations conducted here, and previous experimental results \cite{Jackson2010, Jennings2016} and computational data \cite{Soler2013}, we argue that the experimental enhancement of thermal stability in YibK likely reflects non-equilibrium effects. The persistence of knotted conformations in the denatured ensembles leads to a strong separation of timescales between unfolding (couple of weeks) and unknotting (at least 6 months) \cite{Jennings2016}, and suggests that standard thermal denaturation experiments may not be able to access unfolded and unknotted states, leading to an apparent melting temperatures that does not correspond to a an equilibrium distribution. Overall, our findings indicate that protein knots do not intrinsically enhance thermal stability (or any other thermodynamic equilibrium property), and suggest that research in this topic should focus on non-equilibrium properties instead.

\section{Author contributions}
PFNF designed the research, prepared the figures, and wrote the manuscript. JNCE developed the code and performed all the calculations. All authors analyzed the data. AN, BPT, and MM prepared the \textit{in silico} unknotted version of YibK.

\section{Declaration of interests}
The authors declare no competing interests.

\begin{acknowledgement}
Work supported by UID/04046/2025 Instituto de Biosistemas e Ciências Integrativas Centre Grant from Instituto de Biosistemas e Ciências Integrativas Centre grant from FCT, Portugal. J.N.C.E. acknowledges financial support from FCT, Portugal, through PhD grant SFRH/BD/144345/2019. B.P.T. acknowledges financial support from Fundação Calouste Gulbenkian through the Gulbenkian Novos Talentos Fellowship. This study was also partially supported by the European Union (TWIN2PIPSA - Twinning for Excellence in Biophysics of Protein Interactions and Self-Assembly, GA 101079147). A part of this work was performed on the computational resources of INCD (http://www.incd.pt) funded by FCT and UE under project LISBOA-01-0145-FEDER-022153. Access was granted by FCT through project 2022.26279.CPCA.A0. We thank Sophie Jackson for insightful comments on our manuscript.

\end{acknowledgement}

\bibliography{bibliography}

@string{BBA       = "{B}iochem. {B}iophys. {A}cta"                    }

@string{JAC       = "{J}. {A}ppl. {P}hys."                            }

@string{JCC       = "{J}. {C}omput. {C}hem."                          }

@string{JCP       = "{J}. {C}hem. {P}hys."                            }

@string{JCTC      = "{J}. {C}hem. {T}heory {C}omput."                 }

@string{S         = "{S}cience"                                       }

@article{richardson1977,
  title={$\beta$-Sheet topology and the relatedness of proteins},
  author={Richardson, Jane S},
  journal={Nature},
  volume={268},
  number={5620},
  pages={495--500},
  year={1977},}

@article{taylor2000,
  title={A deeply knotted protein structure and how it might fold},
  author={Taylor, William R},
  journal={Nature},
  volume={406},
  number={6798},
  pages={916--919},
  year={2000},}

@article{Joanna2012,
author = {Joanna I. Sułkowska  and Eric J. Rawdon  and Kenneth C. Millett  and Jose N. Onuchic  and Andrzej Stasiak },
title = {Conservation of complex knotting and slipknotting patterns in proteins},
journal = {Proceedings of the National Academy of Sciences},
volume = {109},
number = {26},
pages = {E1715-E1723},
year = {2012},}

@article{Jackson2020,
  title={Why are there knots in proteins},
  author={Jackson, Sophie E},
  journal={Topol. Geom. Biopolym},
  volume={746},
  pages={129},
  year={2020}}

@article{Grosberg2006,
    author = {Lua, Rhonald C AND Grosberg, Alexander Y},
    journal = {PLOS Computational Biology},
    publisher = {Public Library of Science},
    title = {Statistics of Knots, Geometry of Conformations, and Evolution of Proteins},
    year = {2006},
    month = {05},
    volume = {2},
    pages = {1-8}}

@article{Sulkowska2018,
	author = {Dabrowski-Tumanski, Pawel and Rubach, Pawel and Goundaroulis, Dimos and Dorier, Julien and Su{\l}kowski, Piotr and Millett, Kenneth C and Rawdon, Eric J and Stasiak, Andrzej and Sulkowska, Joanna I},
	issn = {0305-1048},
	journal = {Nucleic Acids Research},
	month = {12},
	number = {D1},
	pages = {D367-D375},
	title = {{KnotProt 2.0: a database of proteins with knots and other entangled structures}},
	volume = {47},
	year = {2018}}

@article{Sulkowska2024,
title = {Everything AlphaFold tells us about protein knots},
journal = {Journal of Molecular Biology},
volume = {436},
number = {19},
pages = {168715},
year = {2024},
issn = {0022-2836},
author = {Agata P. Perlinska and Maciej Sikora and Joanna I. Sulkowska},
}

@article{AlphaFold,
author = {Jumper, John and Evans, Richard and Pritzel, Alexander and Green, Tim and Figurnov, Michael and Ronneberger, Olaf and Tunyasuvunakool, Kathryn and Bates, Russ and {\v Z}{\'\i}dek, Augustin and Potapenko, Anna and Bridgland, Alex and Meyer, Clemens and Kohl, Simon A. A. and Ballard, Andrew J. and Cowie, Andrew and Romera-Paredes, Bernardino and Nikolov, Stanislav and Jain, Rishub and Adler, Jonas and Back, Trevor and Petersen, Stig and Reiman, David and Clancy, Ellen and Zielinski, Michal and Steinegger, Martin and Pacholska, Michalina and Berghammer, Tamas and Bodenstein, Sebastian and Silver, David and Vinyals, Oriol and Senior, Andrew W. and Kavukcuoglu, Koray and Kohli, Pushmeet and Hassabis, Demis},
       journal = {Nature},
	number = {7873},
	pages = {583--589},
	title = {Highly accurate protein structure prediction with AlphaFold},
	volume = {596},
	year = {2021},}

@article{Danny7_1,
title = {Structure, dynamics, and stability of the smallest and most complex 71 protein knot},
journal = {Journal of Biological Chemistry},
volume = {300},
number = {1},
pages = {105553},
year = {2024},
author = {Min-Feng Hsu and Manoj Kumar Sriramoju and Chih-Hsuan Lai and Yun-Ru Chen and Jing-Siou Huang and Tzu-Ping Ko and Kai-Fa Huang and Shang-Te Danny Hsu},
}

@article{Danny2020,
title = {Protein knots provide mechano-resilience to an AAA+ protease-mediated proteolysis with profound ATP energy expenses},
journal = {Biochimica et Biophysica Acta (BBA) - Proteins and Proteomics},
volume = {1868},
number = {2},
pages = {140330},
year = {2020},
author = {Manoj Kumar Sriramoju and Yen Chen and Shang-Te Danny Hsu},
}

@article{Xu2018,
	author = {Xu, Yan and Li, Shixin and Yan, Zengshuai and Luo, Zhen and Ren, Hao and Ge, Baosheng and Huang, Fang and Yue, Tongtao},
	journal = {Biophysical Journal},
	number = {9},
	pages = {1681--1689},
	title = {Stabilizing Effect of Inherent Knots on Proteins Revealed by Molecular Dynamics Simulations},
	volume = {115},
	year = {2018},}

@article{Jackson2017,
author = {Sophie E Jackson and Antonio Suma and Cristian Micheletti},
title = {How to fold intricately: using theory and experiments to unravel the properties of knotted proteins},
journal = {Current Opinion in Structural Biology},
volume = {42},
pages = {6-14},
year = {2017},
}

@article{Soler2013,
    author = {Soler, Miguel A. AND Faísca, Patrícia F. N.},
    journal = {PLOS ONE},
    publisher = {Public Library of Science},
    title = {Effects of Knots on Protein Folding Properties},
    year = {2013},
    volume = {8},
    pages = {1-10},
    number = {9},
}

@article{Soler2014,
       author = {Soler, Miguel A. and Nunes, Ana and Fa{\'\i}sca, Patr{\'\i}cia F. N.},
	journal = {The Journal of Chemical Physics},
	number = {2},
	pages = {025101},
	title = {Effects of knot type in the folding of topologically complex lattice proteins},
	volume = {141},
	year = {2014},}

@article{Joanna2008,
author = {Joanna I. Sułkowska  and Piotr Sułkowski  and P. Szymczak  and Marek Cieplak },
title = {Stabilizing effect of knots on proteins},
journal = {Proceedings of the National Academy of Sciences},
volume = {105},
number = {50},
pages = {19714-19719},
year = {2008},}

@article{Ya-Ming2016,
	author = {Christian, Thomas and Sakaguchi, Reiko and Perlinska, Agata P and Lahoud, Georges and Ito, Takuhiro and Taylor, Erika A and Yokoyama, Shigeyuki and Sulkowska, Joanna I and Hou, Ya-Ming},
	journal = {Nature Structural \& Molecular Biology},
	number = {10},
	pages = {941--948},
	title = {Methyl transfer by substrate signaling from a knotted protein fold},
	volume = {23},
	year = {2016}}

@article{Sara2024,
	author = {Ferreira, Sara G. F. and Sriramoju, Manoj K. and Hsu, Shang-Te Danny and Fa{\'\i}sca, Patr{\'\i}cia F. N. and Machuqueiro, Miguel},
	journal = {Journal of Chemical Information and Modeling},
	number = {17},
	pages = {6827--6837},
	title = {Is There a Functional Role for the Knotted Topology in Protein UCH-L1?},
	volume = {64},
	year = {2024}}

@ARTICLE{Nureki2002,
	author = {Nureki, Osamu and Shirouzu, Mikako and Hashimoto, Kyoko and Ishitani, Ryuichiro and Terada, Takaho and Tamakoshi, Masatada and Oshima, Tairo and Chijimatsu, Masao and Takio, Koji and Vassylyev, Dmitry G. and Shibata, Takehiko and Inoue, Yorinao and Kuramitsu, Seiki and Yokoyama, Shigeyuki},
	title = {An enzyme with a deep trefoil knot for the active-site architecture},
    journal={Acta Crystallographica Section D: Biological Crystallography},
	year = {2002},
	volume = {58},
	number = {7},
	pages = {1129 – 1137},}

@article{Nureki2004,
title = {Deep Knot Structure for Construction of Active Site and Cofactor Binding Site of tRNA Modification Enzyme},
journal = {Structure},
volume = {12},
number = {4},
pages = {593-602},
year = {2004},
author = {Osamu Nureki and Kazunori Watanabe and Shuya Fukai and Ryohei Ishii and Yaeta Endo and Hiroyuki Hori and Shigeyuki Yokoyama},}

@ARTICLE{Jacobs2002,
	author = {Jacobs, Steven A. and Harp, Joel M. and Devarakonda, Srikripa and Kim, Youngchang and Rastinejad, Fraydoon and Khorasanizadeh, Sepideh},
	title = {The active site of the SET domain is constructed on a knot},
    journal={{N}at. {S}truct. {B}iol.},
	year = {2002},
	volume = {9},
	number = {11},
	pages = {833 – 838},}

@article{Virnau2006,
    author = {Virnau, Peter AND Mirny, Leonid A AND Kardar, Mehran},
    journal = {PLOS Computational Biology},
    title = {Intricate Knots in Proteins: Function and Evolution},
    year = {2006},
     volume = {2},
    pages = {1-6},
    number = {9},
}

@article{Faisca2015,
author = {Patrícia F.N. Faísca},
title = {Knotted proteins: A tangled tale of Structural Biology},
journal = {Computational and Structural Biotechnology Journal},
volume = {13},
pages = {459-468},
year = {2015},}

@article{Danny2023,
author = {Shang-Te Danny Hsu},
title = {Folding and functions of knotted proteins},
journal = {Current Opinion in Structural Biology},
volume = {83},
pages = {102709},
year = {2023},
}

@article{Tubiana2024,
title = {Topology in soft and biological matter},
journal = {Physics Reports},
volume = {1075},
pages = {1-137},
year = {2024},
note = {Topology in soft and biological matter},
author = {Luca Tubiana and Gareth P. Alexander and Agnese Barbensi and Dorothy Buck and Julyan H.E. Cartwright and Mateusz Chwastyk and Marek Cieplak and Ivan Coluzza and Simon Čopar and David J. Craik and Marco {Di Stefano} and Ralf Everaers and Patrícia F.N. Faísca and Franco Ferrari and Achille Giacometti and Dimos Goundaroulis and Ellinor Haglund and Ya-Ming Hou and Nevena Ilieva and Sophie E. Jackson and Aleksandre Japaridze and Noam Kaplan and Alexander R. Klotz and Hongbin Li and Christos N. Likos and Emanuele Locatelli and Teresa López-León and Thomas Machon and Cristian Micheletti and Davide Michieletto and Antti Niemi and Wanda Niemyska and Szymon Niewieczerzal and Francesco Nitti and Enzo Orlandini and Samuela Pasquali and Agata P. Perlinska and Rudolf Podgornik and Raffaello Potestio and Nicola M. Pugno and Miha Ravnik and Renzo Ricca and Christian M. Rohwer and Angelo Rosa and Jan Smrek and Anton Souslov and Andrzej Stasiak and Danièle Steer and Joanna Sułkowska and Piotr Sułkowski and De Witt L. Sumners and Carsten Svaneborg and Piotr Szymczak and Thomas Tarenzi and Rui Travasso and Peter Virnau and Dimitris Vlassopoulos and Primož Ziherl and Slobodan Žumer},}

@article{Tsai1999,
title = {The packing density in proteins: standard radii and volumes},
author = {Jerry Tsai and Robin Taylor and Cyrus Chothia and Mark Gerstein},
journal = {Journal of Molecular Biology},
volume = {290},
number = {1},
pages = {253-266},
year = {1999},}

@article{Fernandez2021,
    author  = {Fernández del Río, Beatriz and Rey, Antonio},
    title   = {Behavior of Proteins under Pressure from Experimental Pressure-Dependent Structures},
    journal = {The Journal of Physical Chemistry B},
    volume  = {125},
    number  = {23},
    pages   = {6179-6191},
    year    = {2021},
    }

@article{Larriva2010,
    author  = {Larriva, María and Prieto, Lidia and Bruscolini, Pierpaolo and Rey, Antonio},
    title   = {A simple simulation model can reproduce the thermodynamic folding intermediate of apoflavodoxin},
    journal = {Proteins: Structure, Function, and Bioinformatics},
    volume  = {78},
    number  = {1},
    pages   = {73-82},
    year    = {2010}
}

@article{faisca2019,
author ={Especial, João and Nunes, Ana and Rey, Antonio and Faísca, Patrícia FN},
title  ={Hydrophobic confinement modulates thermal stability and assists knotting in the folding of tangled proteins},
journal  ={Phys. Chem. Chem. Phys.},
year  ={2019},
volume  ={21},
pages  ={11764-11775},
}

@article{Especial2022,
    author  = {Especial, João N. C. and Rey, Antonio and Faísca, Patrícia F. N.},
    title   = {A Note on the Effects of Linear Topology Preservation in Monte Carlo Simulations of Knotted Proteins},
    journal = {Int. J. Mol. Sci.},
    volume  = {23},
    pages   = {13871},
    year    = {2022},
}

@article{Go,
author = {Go, Nobuhiro and Taketomi, Hiroshi},
title = {Studies on protein folding, unfolding and fluctuations by computer simulation IV. Hydrophobic Interactions},
journal = {International Journal of Peptide and Protein Research},
volume = {13},
number = {5},
pages = {447-461},
year = {1979}
}

@article{Yeates2010,
author = {Neil P. King  and Alex W. Jacobitz  and Michael R. Sawaya  and Lukasz Goldschmidt  and Todd O. Yeates },
title = {Structure and folding of a designed knotted protein},
journal = {Proceedings of the National Academy of Sciences},
volume = {107},
number = {48},
pages = {20732-20737},
year = {2010},}

@article{Faisca2012,
    author = {Faísca, Patrícia F. N. and Travasso, Rui D. M. and Parisi, Andrea and Rey, Antonio},
    journal = {PLOS ONE},
    title = {Why Do Protein Folding Rates Correlate with Metrics of Native Topology?},
    year = {2012},
    month = {04},
    volume = {7},
    pages = {1-7},
    number = {4},
}

@article{Danny2019a,
author = {Ya-Chu Chuang and I-Chen Hu and Ping-Chiang Lyu and Shang-Te Danny Hsu},
title = {Untying a Protein Knot by Circular Permutation},
journal = {Journal of Molecular Biology},
volume = {431},
number = {4},
pages = {857-863},
year = {2019},}

@article{Danny2019b,
	author = {Ko, Kuang-Ting and Hu, I-Chen and Huang, Kai-Fa and Lyu, Ping-Chiang and Hsu, Shang-Te Danny},
	journal = {Structure},
	number = {8},
	pages = {1224--1233.e4},
	title = {Untying a Knotted SPOUT RNA Methyltransferase by Circular Permutation Results in a Domain-Swapped Dimer},
	volume = {27},
	year = {2019},}

@article{Jennings2016,
	author = {Capraro, Dominique T. and Jennings, Patricia A.},
	journal = {Biophysical Journal},
	month = {2025/10/12},
	number = {5},
	pages = {1044--1051},
	title = {Untangling the Influence of a Protein Knot on Folding},
	volume = {110},
	year = {2016},}

@article{Jackson2010,
author = {Anna L. Mallam  and Joseph M. Rogers  and Sophie E. Jackson },
title = {Experimental detection of knotted conformations in denatured proteins},
journal = {Proceedings of the National Academy of Sciences},
volume = {107},
pages = {8189-8194},
year = {2010},}

@article{abraham2015,
title = {GROMACS: High performance molecular simulations through multi-level parallelism from laptops to supercomputers},
journal = {SoftwareX},
volume = {1-2},
pages = {19-25},
year = {2015},
author = {Mark James Abraham and Teemu Murtola and Roland Schulz and Szilárd Páll and Jeremy C. Smith and Berk Hess and Erik Lindahl}
}

@article{maier2015,
  title={ff14SB: improving the accuracy of protein side chain and backbone parameters from ff99SB},
  author={Maier, James A and Martinez, Carmenza and Kasavajhala, Koushik and Wickstrom, Lauren and Hauser, Kevin E and Simmerling, Carlos},
  journal=JCTC,
  volume={11},
  number={8},
  pages={3696--3713},
  year={2015},
}

@article{jorgensen1983,
  title={Comparison of simple potential functions for simulating liquid water},
  author={Jorgensen, William L and Chandrasekhar, Jayaraman and Madura, Jeffry D and Impey, Roger W and Klein, Michael L},
  journal=JCP,
  volume={79},
  number={2},
  pages={926--935},
  year={1983},
}

@article{neria1996,
  title={Simulation of activation free energies in molecular systems},
  author={Neria, Eyal and Fischer, Stefan and Karplus, Martin},
  journal=JCP,
  volume={105},
  number={5},
  pages={1902--1921},
  year={1996},
}

@article{darden1993,
title={Particle mesh {E}wald: An {N}log({N}) method for {E}wald sums in
                  large systems},
author={Darden, T. and York, D. and Pedersen, L.},
journal=JCP,
volume={98},
pages={10089--10092},
year={1993}
}

@article{hess2008b,
author={Hess, B.},
title={P-{LINCS}: A Parallel Linear Constraint Solver for
                  Molecular Simulation},
journal=JCTC,
year=2008,
volume=4,
pages={116--122}
}

@article{miyamoto1992,
title={{SETTLE}: {A}n analytical version of the {SHAKE} and {RATTLE}
                  algorithm for rigid water models},
author={Miyamoto, S. and Kollman, P. A.},
journal=JCC,
year={1992},
volume={13},
pages={952--962}
}

@article{bussi2007,
title={Canonical sampling through velocity rescaling},
author={Bussi, G. and Donadio, D. and Parrinello, M.},
journal=JCP,
volume={126},
pages={014101},
year={2007}
}

@article{parrinello1981,
author = {Parrinello, M. and Rahman, A.},
journal = JAC,
month = {dec},
number = {12},
pages = {7182--7190},
title = {{Polymorphic transitions in single crystals: A new molecular
                  dynamics method}},
volume = {52},
year = {1981}
}

@article{Taylor2003,
	author = {Taylor, William R. and Lin, Kuang},
	journal = {Nature},
	number = {6918},
	pages = {25--25},
	title = {Protein knots: A tangled problem},
	volume = {421},
	year = {2003},
}

@article{Joanna_Israelli,
author = {Piejko, Maciej and Niewieczerzal, Szymon and Sulkowska, Joanna I.},
title = {The Folding of Knotted Proteins: Distinguishing the Distinct Behavior of Shallow and Deep Knots},
journal = {Israel Journal of Chemistry},
volume = {60},
number = {7},
pages = {713-724},
year = {2020}}

@Article{faisca2016,
author ="Soler, Miguel A. and Rey, Antonio and Faísca, Patrícia F. N.",
title  ="Steric confinement and enhanced local flexibility assist knotting in simple models of protein folding",
journal  ="Phys. Chem. Chem. Phys.",
year  ="2016",
volume  ="18",
issue  ="38",
pages  ="26391-26403",}

@article{Faisca2010,
year = {2010},
volume = {7},
number = {1},
pages = {016009},
author = {Faísca, Patrícia F N and Travasso, Rui D M and Charters, Tiago and Nunes, Ana and Cieplak, Marek},
title = {The folding of knotted proteins: insights from lattice simulations},
journal = {Physical Biology},}

\end{document}